\documentclass[twocolumn,showpacs,preprintnumbers,amsmath,amssymb]{revtex4}

\usepackage{graphicx}
\usepackage{dcolumn}
\usepackage{bm}

\newcommand{\bra}[1]{\langle#1\rvert}
\newcommand{\ket}[1]{\lvert#1\rangle}
\newcommand{\avr}[1]{\langle#1\rangle}
\renewcommand{\vec}[1]{\bm{\mathrm{#1}}}

\expandafter\ifx\csname
 natexlab\endcsname\relax\fi
 \expandafter\ifx\csname bibnamefont\endcsname\relax
   \def\bibnamefont#1{#1}\fi
 \expandafter\ifx\csname bibfnamefont\endcsname\relax
   \def\bibfnamefont#1{#1}\fi
 \expandafter\ifx\csname citenamefont\endcsname\relax
   \def\citenamefont#1{#1}\fi
 \expandafter\ifx\csname url\endcsname\relax
   \def\url#1{\texttt{#1}}\fi
 \expandafter\ifx\csname
 urlprefix\endcsname\relax\fi
 \providecommand{\bibinfo}[2]{#2}
 \providecommand{\eprint}[2][]{\url{#2}}

\begin{document}

\title{Coherent control of collective spontaneous emission
 in an extended atomic ensemble and quantum storage}

\author{Alexey Kalachev}
\email{kalachev@kfti.knc.ru}
\affiliation{%
Zavoisky Physical-Technical Institute of the Russian Academy of
Sciences, Sibirsky Trakt 10/7, Kazan, 420029, Russia
}%

\author{Stefan Kr\"{o}ll}
\email{Stefan.Kroll@fysik.lth.se}
\affiliation{%
Department of Physics, Lund Institute of Technology (LTH), Box 118,
S-221 00 Lund, Sweden
}%

\date{\today}

\begin{abstract}
Coherent control of collective spontaneous emission in an extended
atomic ensemble resonantly interacting with single-photon wave
packets is analyzed. A scheme for coherent manipulation of
collective atomic states is developed such that superradiant
states of the atomic system can be converted into subradiant ones
and \emph{vice versa}. Possible applications of such a scheme for
optical quantum state storage and single-photon wave packet
shaping are discussed. It is shown that also in the absence of
inhomogeneous broadening of the resonant line,
 single-photon wave
packets with arbitrary pulse shape may be recorded as a subradiant
state and reconstructed even although the duration of the wave
packets is larger than the superradiant life-time. Specifically
the applicability for storing time-bin qubits, which are used in
quantum cryptography is analyzed.
\end{abstract}


\maketitle

\section{\label{sec:level1}Introduction}
Motivated by the development of modern quantum information science
there is significant interest in investigating the interaction of
nonclassical states of light with atomic ensembles. One active
research area is the development of optical quantum memories. Such
devices, which can store and reconstruct quantum states of light,
form a basic ingredient for optical quantum computers, dealing
either with discrete quantum variables (qubits) represented by
single-photon two-mode states of the electromagnetic field or with
continuous quantum variables represented by the quadrature
amplitudes of the field (see reviews \cite{KMNRDM_2005,BL_2005}
respectively). There are experimental demonstrations based on
electromagnetically induced transparency \cite{CMJLKK_2005} for the
discrete variable approach and off-resonant interaction of light
with spin polarized atomic ensembles \cite{JSCFP_2004} for the
continuous variable case. The photon echo based quantum memory
proposal has also attracted considerable attention
\cite{KM_1993,MK_2001,MTH_2003,KTG_2005,NK_2005}. Another cause of
the interest is the possibility of atomic ensemble state
manipulation using atom-photon correlation (see review
\cite{L_2003}). The creation of a robust entanglement of atomic
ensembles using Raman scattering may be especially important for
applications involving long-distance quantum communication
\cite{DLCZ_2001}. Finally, much effort has been directed toward
implementation of controllable sources of nonclassical states of
light such as photon-number or Fock states, which represent an
essential resource for the practical implementation of many ideas
from quantum information. These sources can also be constructed
using quantum storage techniques in optically dense atomic media
\cite{ECAMZL_2004}.

The motivation of all these investigations is that photons can
interact much more strongly with ensembles containing a large number
of atoms than with individual atoms. In addition there can be a
further collective enhancement taking place because the absorption,
emission or coherent scattering processes employed in several of
these implementation schemes involve collective atomic modes. On the
other hand, whenever the number of atoms is large, the contribution
from the collective spontaneous emission
\cite{D_1954,GH_1982,BEMST_1996} to the evolution of an atomic-field
system can also be considerable and may need to be taken into
account. Therefore the investigation of this phenomenon in various
situations when light interacts with atomic ensembles is of
interest.

In this paper we investigate the possibilities of coherent control
of collective spontaneous emission in an extended atomic ensemble
interacting with a single-photon wave packet and vacuum modes of the
electromagnetic field. Spontaneous emission is normally considered
as a noise, which should be inhibited. Various schemes for
inhibiting or reducing spontaneous emission noise for the single
atom case have been proposed and investigated. For example in
\cite{ASW_2001} excitation by $2\pi$-pulses on an auxiliary
transition is used to change the phase of the excited state wave
function and in this way inhibit the spontaneous decay. In an atomic
ensemble confined to a volume with dimensions small compared to the
wavelength of the emitted radiation the problem is reduced to the
creation of antisymmetric subradiant states
\cite{D_1954,MW_1995,DVB_1996}, which are also interesting as they
may be a basis for forming a decoherence free subspace
\cite{LW_2003} for the excited ensemble. In extended atomic systems,
which is a much more frequent experimental situation, the collective
spontaneous emission is characterized by a sharp directedness in
space and occurs only in a small fraction of all radiation modes.
Therefore it is possible to implement coherent control of collective
atomic states using coherent excitation through other non-collective
modes. Such a manipulation of collective states may be used for
quantum memories in the regime of optical subradiance inhibiting the
normal spontaneous decay, an idea which was briefly outlined in
\cite{KS_2005}.

The paper is organized as follows. In Sec.~II, we introduce a
physical model of interaction between an extended many atom system
and a single-photon wave packet and present an analytical
description of the superradiant forward scattering. In Sec.~III, we
present a scheme for coherent manipulation of collective atomic
states that can transform superradiant states into subradiant ones
and \emph{vice versa}. In Sec.~IV, we discuss possible applications
of such a scheme for the creation of optical quantum memory devices
and controlled sources of non-classical light states. In Sec.~V, we
propose some specific experimental implementations that may be used
to observe these effects and some possible methods for subradiant
state preparation.

\section{Physical model}
Consider a system of $N\gg 1$ identical two-level atoms, with
positions $\vec{r}_j$ ($j=1,\ldots,N$) and resonance frequency
$\omega_0$, interacting among themselves and with the external world
only through the electromagnetic field. Let us denote the ground and
excited states of $j$th atom by $\ket{0_j}$ and $\ket{1_j}$. The
Hamiltonian of the system, in the interaction picture and
rotating-wave approximation, reads
\begin{equation}\label{Ham}
H=\sum_{j,\vec{k},s}\hbar g_{\vec{k},s}^\ast b_j^\dag
a_{\vec{k},s}{\,e}^{i\vec{k}\cdot\vec{r}_j}
{\,e}^{i(\omega_0-\omega)t}+\text{H.c.}
\end{equation}
Here
\begin{equation}
g_{\vec{k},s}=\frac{i}{\hbar}\left(\frac{\hbar\omega}{2\varepsilon_0
V}\right)^{1/2}(\vec{d}\cdot\vec{\varepsilon}_{\vec{k},s})
\end{equation}
is the atom-field coupling constant, $b_j=\ket{0_j}\bra{1_j}$ is the
atomic transition operator, $a_{\vec{k},s}$ is the photon
annihilation operator in the radiation field mode with the frequency
$\omega=kc$ and polarization unit vector
$\vec{\varepsilon}_{\vec{k},s}$ ($s=1,2$), $V$ is the quantization
volume of the radiation field (we take $V$ much larger than the
volume of the atomic system), $\vec{d}$ is the dipole moment of the
atomic transition. For the sake of simplicity we assume that the
vectors $\vec{\varepsilon}_{\vec{k},s}$ and $\vec{d}$ are real.

To describe the interaction of the field with the atoms we use the approach
\cite{BL_1975,BB_1975}. First, it is convenient to assume that\\
(i) the atomic system has a shape of a parallelepiped with the
dimensions $L_\alpha$, ($\alpha=x,y,z$) and the atoms are placed
in a regular cubic lattice, so that $N=N_xN_yN_z$ and $d=L_\alpha
N_\alpha^{-1}$ is the interatomic distance,\\ (ii) the dimensions
$L_\alpha$ are much larger, but the interatomic distance $d$ is
much smaller, than the
wavelength $\lambda=2\pi c\omega_0^{-1}$ of the atomic transition,\\
(iii) the center of the system lies in the origin of the reference
frame. Then we can define the following collective atomic operators:
\begin{equation}
R_{\vec{q}}=\sum_{j=1}^N b_j{\,e}^{- i\vec{q}\cdot\vec{r}_j},
\end{equation}
where ${q}_\alpha=2\pi n_\alpha L_\alpha^{-1}$,
$n_\alpha=0,1,\ldots,N_\alpha-1$. Hamiltonian (\ref{Ham}) can now be
expressed in terms of the collective operators obtaining
\begin{equation}\label{H}
H=\sum_{\vec{q},\vec{k},s}\hbar g_{\vec{k},s}^\ast
R^{\dag}_{\vec{q}}
a_{\vec{k},s}\phi^\ast(\vec{q}-\vec{k})
{\,e}^{i(\omega_0-\omega)t}+\text{H.c.},
\end{equation}
where
$\phi(\vec{x})={N}^{-1}\sum_{j}{\exp}(i\vec{x}\cdot\vec{r}_j)$ is
the diffraction function. The presence of this function in the
Hamiltonian underlines the fact that each atomic mode $\vec{q}$ is
coupled only to modes $\vec{k}$ lying in a diffraction angle
around $\vec{q}$ \cite{BL_1975}.

We are interested in the interaction of the atomic system with a
single-photon wave packet. Therefore we have the following general
form of the state of the system at the initial time $t=0$
\begin{equation}\label{psi_0}
\ket{\psi(0)}=\sum_{\vec{k},s}f_{\vec{k},s}(0)\ket{0}\ket{1_{\vec{k},s}}+
\sum_{\vec{q}}c_{\vec{q}}(0)\ket{1_{\vec{q}}}\ket{\text{vac}}
\end{equation}
with normalization condition
\begin{equation}
\sum_{\vec{k},s}|f_{\vec{k},s}(t)|^2+\sum_{\vec{q}}|c_{\vec{q}}(t)|^2=1,
\end{equation}
where $\ket{0}=\ket{0_1,0_2,\ldots,0_N}$ is the ground state of the
atomic system, $\ket{\text{vac}}$ is the vacuum state of the
radiation field,
$\ket{1_{\vec{k},s}}=a^\dag_{\vec{k},s}\ket{\text{vac}}$ and
$\ket{1_{\vec{q}}}={N}^{-1/2}R^{\dag}_{\vec{q}}\ket{0}$.

Now, we assume that\\ (iv) a single-photon wave packet propagates in
the $z$-direction, the excitation volume may be approximated by a
cylinder with the cross section $S$ and the length $L_z$, and the
wave front of the packet is planar inside the excitation volume,\\
(v) the Fresnel number of the excitation volume
$F=S(L_z\lambda)^{-1}\approx 1$, so that the collective interaction
of the whole ensemble of atoms with the quantum field takes place
only for longitudinal modes $\vec{q}$,\\
(vi) the duration of the single-photon wave packet as well as that
of collective spontaneous emission are much greater than
cooperative time $\tau_c$ \cite{AC_1970}, which in turn is much
greater than the correlation time of the vacuum reservoir
$\tau_E=L_zc^{-1}$ (Born-Markov condition). Here
\begin{equation}
\tau_c=\left(\frac{8\pi T_1 SL}{3\lambda^2 c N}\right)^{1/2},
\end{equation}
where $T_1$ is the inverse of the Einstein $A$ coefficient. With
these assumptions the interaction of the atomic system with the
electromagnetic field may be described in a one-mode approximation
with respect to the atomic system and a one-dimensional
approximation with respect to the field. Substituting
Eqs.~(\ref{H}) and (\ref{psi_0}) in the Schr\"{o}dinger equation
and omitting the index $\vec{q}$ for the single atomic mode we
obtain
\begin{eqnarray}
\frac{\partial
f_{\vec{k},s}(t)}{\partial t}&=&-ig_{\vec{k},s}\sqrt{N}\phi(\vec{q}-\vec{k})
c(t){\,e}^{-i(\omega_0-\omega)t},\label{Eq1}\\
\frac{\partial c(t)}{\partial t}&=&-i\sqrt{N}\nonumber\\&
&\times\sum_{\vec{k},s}g^\ast_{\vec{k},s}\phi^\ast(\vec{q}-\vec{k})
f_{\vec{k},s}(t){\,e}^{i(\omega_0-\omega)t}.\label{Eq2}
\end{eqnarray}

Now it is convenient to define a single-photon dimensionless photon
density at the origin of the reference frame ($\vec{r}=0$). For the
incoming wave packet we have
\begin{equation}
F_\text{in}(t)=\left(\frac{L_xL_yL_z}{L^3}\right)^{1/2}\sum_{\vec{k},s}
f_{\vec{k},s}(0){\,e}^{i(\omega_0-\omega)t},
\end{equation}
and for the emitted radiation we have the analogous equation with
$F_\text{in}(t)$ and $f_{\vec{k},s}(0)$ replaced by $F(t)$ and
$f_{\vec{k},s}(t)$, respectively. Then the solution of
Eqs.~(\ref{Eq1}) and (\ref{Eq2}) may be written as (see Appendix
\ref{app:A})
\begin{eqnarray}
c(t)&=&c(0){\,e}^{- t/2\tau_R}\nonumber\\&
&-\frac{1}{\sqrt{\tau_R\tau_E}}\int_0^\infty d\tau
F_\text{in}(t-\tau){\,e}^{-\tau/2\tau_R},\label{Solution_0a}\\
F(t)&=&F_\text{in}(t)+\sqrt{\frac{\tau_E}{\tau_R}}c(t),\label{Solution_0b}
\end{eqnarray}
where $\tau_R$ is the superradiant life-time:
\begin{equation}\label{tau_R}
\frac{1}{\tau_R}=\frac{N\mu}{T_1},\quad
\frac{1}{T_1}=\frac{1}{4\pi\varepsilon_0}\frac{4d^2\omega_0^3}{3\hbar
c^3},
\end{equation}
and $\mu=3\lambda^2(8\pi S)^{-1}$ is a geometrical factor, which is
approximately equal to the ratio of the diffraction solid angle of
collective emission to $4\pi$. If we consider the case when $c(0)=0$
and substitute Eq.~(\ref{Solution_0a}) into (\ref{Solution_0b}), we
obtain a solution for superradiant resonant forward scattering of
photons, which is well known in the theory of propagation of
coherent pulses through a resonant medium and especially in the
coherent-path model of nuclear resonant scattering of $\gamma$
quanta \cite{H_1997}:
\begin{equation}\label{Fscat}
F(t)=F_\text{in}(t)-\frac{1}{\tau_R}\int_0^\infty d\tau
F_\text{in}(t-\tau){\,e}^{-\tau/2\tau_R}.
\end{equation}
Under the approximations mentioned earlier, Eq.~(\ref{Fscat})
describes the simplest regime of superradiant forward scattering,
when the incoming field is scattered only once. Considering
resonant propagation, e.g. of small area pulses as in
\cite{C_1970}, this corresponds to samples with $\alpha L$ not far
above 1. It may be noted that the collective superradiant atomic
state $\ket{1_{\vec{q}}}$ is known in the theory of nuclear
resonant scattering as a nuclear exciton (see, for example,
\cite{S_1999,OH_2005} and references therein). The principal point
here is the indistinguishability of atoms in a macroscopic sample
with respect to photon absorption and emission into the
longitudinal modes of the electromagnetic field, which leads to
enhancement of atomic-field interaction and directedness of
collective spontaneous emission \cite{SFOW_2006}. Since the photon
propagation time $\tau_E$ is much less than the cooperative time
$\tau_c=\sqrt{\tau_R\tau_E}$, which in turn is much less than the
duration of incoming and scattered photons, propagation effects
may be neglected. In fact, from the view point of propagating
fields, the sample looks like a scattering center, which is
characterized by an enhanced cross-section of the resonant
transition with respect to longitudinal modes. But the interaction
of the atomic system with transverse modes of electromagnetic
field is not collective and these modes do not see an enhanced
cross section. This is of key importance for the preparation of
subradiant states in an extended atomic system, which will be
considered in the next section.

\section{Superradiant and subradiant states of an extended atomic ensemble}
In his basic paper \cite{D_1954} Dicke considered the two regimes of
collective spontaneous emission of photons: superradiance and
subradiance, which result from the constructive and destructive
interference of atomic states, respectively. In the first case, a
system of inverted atoms undergoes the spontaneous transition to the
ground state for a time inversely proportional to the number of
atoms, while in the second case the rate of collective spontaneous
emission, on the contrary, decreases compared to the rate of
spontaneous emission of single atoms. In the ideal case of a
localized system confined to a volume with dimensions small compared
to the wavelength of the emitted radiation, the rate of collective
spontaneous emission of photons is equal to zero if the atoms are in
an antisymmetric collective state. If an atomic system is extended,
the rate of collective spontaneous emission can be suppressed only
for a few collective modes. In particular, one-mode subradiance can
be observed in samples having a shape of a cylinder with proportions
defined by the Fresnel number $F\approx 1$. It is well known that
such a one-mode model of collective spontaneous emission is
equivalent to the Dicke model \cite{BSH_1971}, except that the rate
of spontaneous emission decreases by the geometrical factor $\mu$.
If we redefine atomic states $\ket{1_j}$ multiplying them by phase a
factor $\exp(i\vec{q}\cdot\vec{r}_j)$ and denote a collective
one-mode atomic state $\ket{n_{\vec{q}}}$ corresponding to the
excitation of $n$ atoms as $\ket{n}$, then
\begin{eqnarray}
\ket{1}&=&\frac{1}{\sqrt{N}}R^{\dag}\ket{0}=
\frac{1}{\sqrt{N}}\sum_{i}\ket{0,\ldots,1_i,\ldots,0},\label{1N}\\
\ket{2}&=&\frac{1}{\sqrt{2N(N-1)}}(R^{\dag})^2\ket{0}\nonumber\\
&=&\sqrt{\frac{2}{N(N-1)}}
\sum_{j>i}\ket{0,\ldots,1_i,\ldots,1_j,\ldots,0},\label{2N}
\end{eqnarray}
etc., where $R=\sum_{j=1}^N b_j$. The rate of collective spontaneous
emission of photons ${\mu}{T_1}^{-1}|\bra{n-1}R\ket{n}|^2$ is
increased during excitation of the medium (for $n < \frac{N}{2}$).
The spontaneous emission of a photon from the state $\ket{1}$ occurs
with the rate $N\mu T_1^{-1}$ (see Eq.~(\ref{tau_R})), from the
state $\ket{2}$ with the rate $2(N-1)\mu T_1^{-1}$, etc.

\begin{figure}
\includegraphics[width=5.6cm]{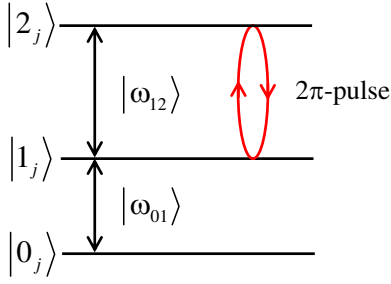}
\caption{\label{fig:levels} (Color online) Scheme of the working
atomic levels, which are used for preparing subradiant collective
states.}
\end{figure}

We assume now that atoms have the additional level $\ket{2_j}$ and
the transition frequency $\omega_{12}$ differs from $\omega_{01}$
(Fig.~\ref{fig:levels}). Then, irradiation by a short coherent
$2\pi$ pulse at the frequency $\omega_{12}$ can almost instantly
change the phase of the states $\ket{1_j}$, thereby changing the
phase of the collective state $\ket{1}$ to the opposite one, while
the phase of the state $\ket{2}$ remains unchanged.

Let us divide a sample into two parts, $A$ and $B$, each containing
$\frac{N}{2}$ atoms. It is convenient to represent superradiant
states (\ref{1N}) and (\ref{2N}) in the form
\begin{eqnarray}
\ket{1_{A+B}}&=&\frac{1}{\sqrt{2}}
\big(\ket{1_{A},0_{B}}+\ket{0_{A},1_{B}}\big),\label{1AB}\\
\ket{2_{A+B}}&=&\frac{1}{2\sqrt{N-1}}[\sqrt{N-2}
(\ket{2_A,0_B}+\ket{0_A,2_B})\nonumber\\&
&+\sqrt{2N}\ket{1_A,1_B}].\label{2AB}
\end{eqnarray}
Upon irradiation of atoms in, for example, part B, by a coherent
$2\pi$-pulse the atomic system passes from states (\ref{1AB}) and
(\ref{2AB}) to the antisymmetric single-mode subradiant states
\begin{eqnarray}
\ket{1_{A-B}}&=&\frac{1}{\sqrt{2}}(\ket{1_A,0_B}-\ket{0_A,1_B}),\label{1A_B}\\
\ket{2_{A-B}}&=&\frac{1}{2\sqrt{N-1}}[
\sqrt{N-2}(\ket{2_A,0_B}+\ket{0_A,2_B})\nonumber\\& &
-\sqrt{2N}\ket{1_A,1_B}].\label{2A_B}
\end{eqnarray}
The rate of photon emission from state (\ref{1A_B}) is identically
zero, while that from state (\ref{2A_B}) is $2(N-1)^{-1}\mu
T_1^{-1}$, i.e., goes to zero for large $N$.

Now assume that the atomic system in state (\ref{1A_B}) absorbs a
photon and passes to the state
\begin{gather}
\ket{2'_{A-B}}=\frac{1}{\sqrt{2}}(\ket{2_A,0_B}-\ket{0_A,2_B}).\label{2prime}
\end{gather}
The rate of photon emission in this case is $(N-2)\mu T_1^{-1}$,
i.e., in fact essentially the same as the superradiant spontaneous
decay from the state $\ket{1}$ at large $N$. To convert the state
in Eq.~(\ref{2prime}) into a subradiant state, it is necessary to
irradiate half of the atoms, whose phase was changed after
absorption of the first photon, and half atoms, whose phase was
not changed, by additional $2\pi$-pulses. As a result, the atomic
system will be divided into four parts, which we denote $A$, $B$,
$C$, and $D$, and the state of the system will take the form
\begin{gather}
\ket{2_{A-B+C-D}}=\frac{1}{\sqrt{2}}
(\ket{2_{A-B},0_{C+D}}-\ket{0_{A+B},2_{C-D}}).
\end{gather}
The rate of collective spontaneous emission from this state is equal
to $4(N-2)^{-1}\mu T_1^{-1}$.

Therefore, by irradiating the system of atoms by $2\pi$ pulses
through auxiliary transition after each absorption of a photon, we
can transfer it into the next higher excited subradiant state, where
the spontaneous transition from this state to the previous
subradiant state will be forbidden. When transfering the atomic
system into subradiant states repeatedly the action of the $2\pi$
pulses can be described using Hadamard matrixes
\begin{equation}
H_2=\left[%
\begin{array}{cc}
  1 & 1 \\
  1 & -1 \\
\end{array}%
\right],\quad
H_{2k}=\left[%
\begin{array}{cc}
  H_k & H_k \\
  H_k & -H_k \\
\end{array}%
\right],
\end{equation}
where $k=2,4,8,$ etc. The scheme for the preparation of different
subradiant states can be seen directly from the matrix where each
column correspond to one specific spatial part and, with the
exception of the first row, each row shows the phase of all the
spatial parts for a specific subradiant state. Transformation from
one row to another is accomplished by phase shift $\ket{1_j}\to
-\ket{1_j}$ for half of the atoms. The size $2k$ is equal to the
number of the spatial parts and the total number of orthogonal
subradiant states which can be prepare in such a way is $2k-1$.

\section{Quantum memory applications and single-photon pulse shaping}

Now, consider the writing and reading of a single-photon wave packet
with a duration $T$ satisfying the condition $\tau_R\ll T$. The
writing process will be divided into several time intervals, the
duration of which is of the order of $\tau_R$ and much less than
that of the single-photon wave packet $T$. Between the intervals,
the atomic system is subjected to $2\pi$ pulses, which transfer it
to different subradiant states. The duration of these pulses $t_p$
should be short in comparison with $\tau_R$. The probability of the
failure of such a transformation is (see Appendix \ref{app:B})
\begin{equation}
p_\text{failure}=|c_0|^2\left(\frac{g_a\Omega_R}
{g_a^2+(\Omega_R/2)^2}\right)^2,
\end{equation}
where $|c_0|^2$ is the probability that the atomic system is in the
ground state and the field in a single-photon state, at the
beginning of the transformation, $\Omega_R$ is the Rabi frequency of
the coherent pulse and $g_a$ is the atom-field coupling constant on
the transition $\ket{0_j}-\ket{1_j}$. Therefore if $\Omega_R\gg
g_a$, $p_\text{failure}$ becomes close to 0.

\begin{figure}
\includegraphics[width=8.6cm]{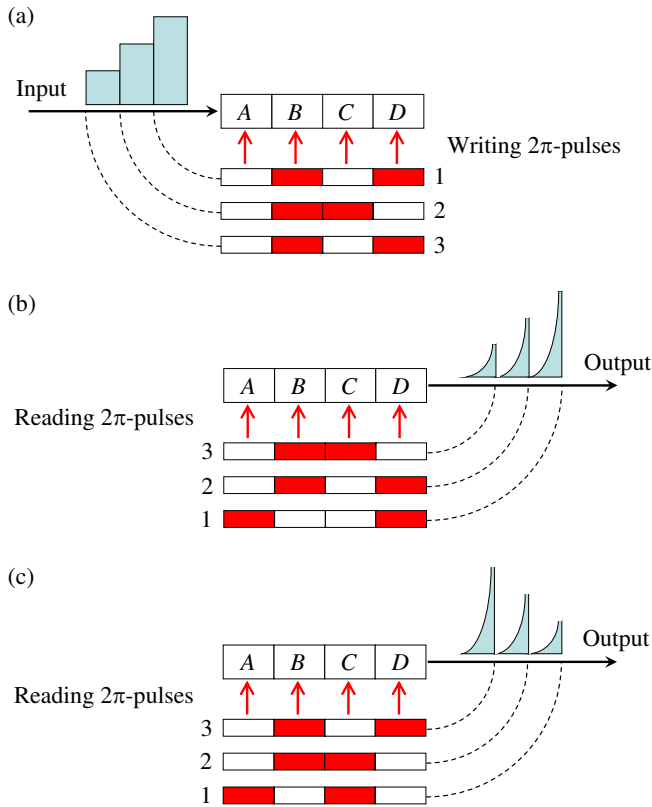}
\caption{\label{fig:scheme} (Color online) Geometry and the
temporal order of the action of $2\pi$-pulses preparing the
subradiant states of an extended medium in the case of four
spatial parts and three possible subradiant states. A filled area
indicates the action of a $2\pi$-pulse in the corresponding
spatial domain: (a) upon writing the quantum state of the
single-photon wave packet, (b) upon read-out without time
reversal, and (c) upon read-out with time reversal.}
\end{figure}

The probability amplitude of photon absorption in each time interval
is proportional to the wave-function amplitude of the photon in
accordance with Eq.~(\ref{Solution_0a}). Therefore, at the end of
the writing process, the state of the atomic system will be
determined by a superposition of orthogonal subradiant states, each
corresponding to that the photon was absorbed in the corresponding
time interval with a weight proportional to the single photon
wave-function amplitude. If now the atomic system is subjected to a
sequence of $2\pi$ pulses, which convert it back from the subradiant
states to the superradiant states, the single-photon wave packet
will be reconstructed, since the probability amplitude of a
spontaneous photon emission is proportional to the amplitude of the
excited atomic state at the start of the corresponding time
intervals in accordance with Eq.~(\ref{Solution_0b}). The simple
example of single-photon writing and reconstruction is shown in
Fig.~\ref{fig:scheme}.
In this case the atomic system is divided into four spatial parts so
that the preparation of three orthogonal subradiant states is
possible, which corresponds to the following Hadamard matrix
\begin{equation}\label{Had1}
H_4=\left(%
\begin{array}{cccc}
  1 & 1 & 1 & 1 \\
  1 & -1 & 1 & -1 \\
  1 & 1 & -1 & -1 \\
  1 & -1 & -1 & 1 \\
\end{array}%
\right).
\end{equation}
Let $c_n$ be the probability amplitude that a photon has been
captured during the $n$-th time interval. It is convenient to write
the photon state in the following form
\[\ket{F}=\sum_n f_n\ket{1_n}, \]
where $\ket{1_n}$ is an elementary single-photon wave packet
propagating through the sample during the $n$-th time interval and
$f_n$ is the probability amplitude to find a photon in the state
$\ket{1_n}$. $\sum_n|c_n|^2=1$ and in the ideal case $c_n=f_n$
(within a phase factor insensitive to $n$). Then the evolution of
the atomic-field system during the application of $2\pi$-pulses in
accordance with the scheme (\ref{Had1}) may be described as follows
\begin{widetext}
\begin{multline*}
\ket{0}(f_1\ket{1_1}+f_2\ket{1_2}+f_3\ket{1_3})\to\\
c_1\ket{1_{A-B+C-D}}\ket{\text{vac}}+\ket{0}(f_2\ket{1_2}+f_3\ket{1_3})\to
(c_1\ket{1_{A+B-C-D}}+c_2\ket{1_{A-B-C+D}})
\ket{\text{vac}}+\ket{0}f_3\ket{1_3}\to\\
(c_1\ket{1_{A-B-C+D}}+c_2\ket{1_{A+B-C-D}}+c_3\ket{1_{A-B+C-D}})
\ket{\text{vac}}.
\end{multline*}
\end{widetext}
The first pulse acts on parts $BD$, the second on $BC$, and the
third on $BD$ again [Fig.~\ref{fig:scheme}(a)]. Now the shape of a
single-photon wave packet is stored as the set of amplitudes $c_n$.
In order to reconstruct the single-photon state we apply three
$2\pi$-pulses, which act on parts $AD$, $BD$, and $BC$, respectively
[Fig.~\ref{fig:scheme}(b)]. It should be noted that after the action
of the first read-out pulse the phase of the emitted wave-packet
becomes the same as for the incoming one. Let $e_n$ be the
probability amplitude that the photon has been emitted between the
application of the $n$-th and $(n+1)$-th read-out pulses. Then the
evolution of the atomic-field system upon the reading of information
reads
\begin{widetext}
\begin{multline*}
(c_1\ket{1_{A-B-C+D}}+c_2\ket{1_{A+B-C-D}}+c_3\ket{1_{A-B+C-D}})
\ket{\text{vac}}\to\\
\ket{0}e_1\ket{1_1}+(c_2\ket{1_{-A+B-C+D}}+c_3\ket{1_{-A-B+C+D}})
\ket{\text{vac}}\to
\ket{0}(e_1\ket{1_1}+e_2\ket{1_2})+c_3\ket{1_{-A+B+C-D}}
\ket{\text{vac}}\to\\
\ket{0}(e_1\ket{1_1}+e_2\ket{1_2}+e_3\ket{1_3}),
\end{multline*}
\end{widetext}
where $e_n=f_n$ in the ideal case. Moreover, by rearranging the
coherent $2\pi$ pulses at the reading stage, we can permute the
superradiant states and form a preassigned shape of the emitted
single-photon wave packet. For example, the application of reading
$2\pi$ pulses to the parts $AC$, $BC$ and $BD$ leads to the reversal
of a single photon wave packet [see Fig.~\ref{fig:scheme}(c)].
Besides, it is possible to modulate the phase of the emitted photon
at each elementary act of read-out by applying $2\pi$-pulses to
spatial regions, which are complimentary to those used for the
writing.

The number of subradiant states used for the storage of a single
photon wave packet determines the time resolution of the quantum
memory. The reconstructed pulse shape in general case is different
from the original one, but the probability distribution within the
time intervals between $2\pi$-pulses is the same. Therefore, such a
quantum memory may be useful, for example, for a quantum
communication using time-bin qubits or, more generally, time-bin
qudits. The procedure described here for the case of a single photon
may be extended in a straightforward way in order to write and
reconstruct a sequence of photons. In this case it is necessary to
increase the total number of orthogonal subradiant states and
corresponding number of spatial parts as was described in Sec.~III.
By rearranging the read-out pulses we can form not only a
preassigned shape of the emitted single-photon wave packets, but
also a preassigned sequence of them.

\begin{figure}
\includegraphics[width=8.6cm]{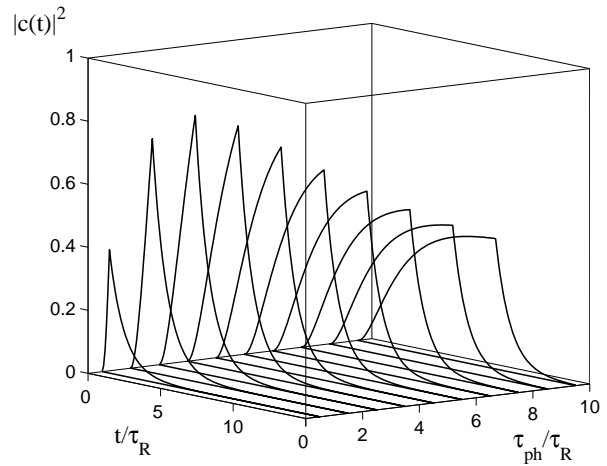}
\caption{\label{fig:population} The population of the excited atomic
state $|c(t)|^2$ as a function of time $t$ and duration of
rectangular incoming wave packet $\tau_{ph}$.}
\end{figure}

In order to estimate the efficiency of the writing and reading
processes consider the solution (\ref{Solution_0a}) for the case
when all atoms are in the ground state at time $t=0$ and the
single-photon wave packet has a quasi-rectangular form with a
duration $\tau_\text{ph}$ and amplitude
$F_\text{in}(t)=\sqrt{\tau_E/\tau_\text{ph}}$. The term
'quasi-rectangular form' means that the leading and trailing
fronts of the pulse are not shorter than $\tau_E$. (If this would
not be the case the one-mode approximation becomes invalid.) Then
from Eq.~(\ref{Solution_0a}) we obtain
\begin{equation}\label{Solution_1a}
c(t)=\left\{%
\begin{array}{ll}
 2\sqrt{\dfrac{\tau_R}{\tau_\text{ph}}}
 \left[\exp\left(-\dfrac{t}{2\tau_R}\right)-1 \right],
 & \text{$t\leq\tau_\text{ph}$,} \\
 c(\tau_\text{ph})\exp\left(-\dfrac{t-\tau_\text{ph}}{2\tau_R}\right),
 & \text{$t>\tau_\text{ph}$.} \\
\end{array}%
\right.
\end{equation}
The maximum of $|c(t)|$ is equal to 0.9 and is attained when
$\tau_\text{ph}\approx 2.5\tau_{R}$.  The population of the excited
atomic state $|c(t)|^2$ as a function of time for the case when the
incoming wave packet has a rectangular shape is shown in
Fig.~\ref{fig:population}.
Thus,  the efficiency of the writing process for a wave packet of
duration $T\gg\tau_R$ is equal to 81\% when the $2\pi$-pulses
separated by $2.5\tau_R$, which is the optimal value. Since the
duration of the elementary read-out act is also equal to
$2.5\tau_R$, the efficiency of read-out of information is equal to
$1-\exp(-2.5)$, i.e., 91\%. Therefore the total efficiency of the
scheme proves to be about 75\%. It is also possible to find pulse
shapes that will enable storage and recall with a similar overall
efficiency of 75\% in a self-consistent regime, where the output
pulse has the same shape as the input pulse.

However, if each pulse only need to be stored and read once the
memory can be operated in a nearly 100\% efficient mode. In
accordance with Eq.~(\ref{Solution_0a}) the probability of
absorption is maximal for a propagating pulse like
\begin{equation}\label{reverse exp}
F_\text{in}(t)=\left\{%
\begin{array}{ll}
    \sqrt{\dfrac{\tau_E}{\tau_R}}
    \exp\left(\dfrac{t-t_\text{end}}{2\tau_R}\right),
    & \text{$t\leq t_\text{end}$,} \\
    0, & \text{$t>t_\text{end}$,} \\
\end{array}%
\right.
\end{equation}
where $t_\text{end}\gg\tau_R$. In this case we have
\begin{equation}\label{Solution_2a}
c(t)=\left\{%
\begin{array}{ll}
    -\exp\left(\dfrac{t-t_\text{end}}{2\tau_R}\right),
    & \text{$t\leq t_\text{end}$,} \\
    -\exp\left(-\dfrac{t-t_\text{end}}{2\tau_R}\right),
    & \text{$t>t_\text{end}$.} \\
\end{array}%
\right.
\end{equation}
Therefore, $|c(t)|^2=1$ at the end of the pulse $t=t_\text{end}$.
Now, if we apply a $2\pi$-pulse to half the atoms at this time, a
subradiant state is created with unit probability. Upon read-out we
obtain (with unit probability) the pulse shape
\begin{equation}\label{usual exp}
F(t)=\sqrt{\frac{\tau_E}{\tau_R}}
\exp\left(-\frac{t-t_\text{read}}{2\tau_R}\right),\quad
t\geq t_\text{read},
\end{equation}
corresponding to spontaneous emission after the application of a
short read-out pulse at the time $t=t_\text{read}$ (the phase of the
emitted photon can be changed by $\pi$ at the read-out as was
considered above). Thus, nearly unit efficiency quantum storage with
time reversal is possible for wave-packets of the form in
Eq.~(\ref{reverse exp}). Such a regime may be useful, for example,
for a long-distance quantum communication using quantum repeaters
\cite{BDCZ_1998}, when the qubits are only stored and recalled once
before being measured. Assuming that faint laser pulses are used for
carrying the information \cite{GRTZ_2002}, the preparation of a
time-bin qubit, which is a superposition of well separated
exponential wave packets like (\ref{reverse exp}), is not a
challenging experimental task. Each of the wave-packets may be
recorded (reconstructed) using only one transformation to (from) a
subradiant state.

In conclusion of this section, it should be noted that the simple
model of quantum storage with time reversal may be generalized to
include propagation effects. In this case the shape of pulses to be
stored should be determined by the time-reversed response function
of a sample. As a result, both recorded and reconstructed
wave-packets prove to be modulated in time, but the unit efficiency
of quantum storage is maintained.

\section{Implementing the scheme in a solid state medium}

Solids and especially rare-earth-ion-doped crystals are attractive
materials for optical and quantum optical memories. At cryogenic
temperatures optical transitions of rare-earth-ion-doped crystals
have very narrow homogeneous lines, which correspond to long phase
relaxation of the optical transitions, up to milliseconds
\cite{M_2002}. But observation of collective spontaneous emission
in solids as well as related effects such as superradiant forward
scattering considered above is a nontrivial experimental problem.
Since inhomogeneous broadening of optical transitions is usually
several GHz and oscillator strengths are small, it is in order to
fulfill the condition $\tau_R\ll T_2^\ast$, necessary to use very
high densities of impurities and perform observations in the ps
regime or to use frequency selective excitation of the
inhomogeneous profile, which significantly limits the superradiant
effect. From this point the technique of preparing of narrow
absorbing peaks on a non-absorbing background, i.e. isolated
spectral features corresponding to a group of ions absorbing at a
specific frequency, in rare-earth-ion-doped crystals
\cite{PSM_2000,SPMK_2000,NROCK_2002,SLLG_2003,NRKKS_2004,RNKKS_2005}
can be very useful. Such specific structures can be created as
follows. First, spectral pits, i.e. wide frequency intervals
within the inhomogeneous absorption profile that are completely
empty of all absorption, are created using hole-burning
techniques. The maximum width of the pits is determined by the
hyperfine level separations \cite{NRKKS_2004}. Then narrow peaks
of absorption can be created by pumping ions absorbing within a
narrow spectral interval back into the emptied region. It is
possible to select only a subset of ions absorbing on a transition
between specific hyperfine levels in the ground and excited states
\cite{NRKKS_2004}. The peaks can have a width of the order of the
homogeneous linewidth, if a laser with a sufficiently narrow
linewidth is used for the preparation.

We now turn to a specific example of how to realize quantum state
storage under the optical subradiance regime, using the
${}^3H_4(1)-{}^1D_2(1)$ transition of $\text{Pr}^{3+}$ ions in
$\text{Y}_2\text{SiO}_5$. The optical properties of this crystal
has been studied extensively and the preparation of narrow
absorption peaks have already been demonstrated on this transition
in experiments mentioned above. The ground state is split into
three doubly degenerate states, which are usually labeled
$\ket{\pm 1/2}$, $\ket{\pm 3/2}$, and $\ket{\pm 5/2}$, separated
by 10.2 and 17.3 MHz \cite{HCVW_1993,ECM_1995} for ions in site 1.
A narrow absorption peak with a line-width of less than
$\Gamma_{\text{inh}}=100$~kHz is usually created in one of these
states on a $10-15$~MHz wide pit in the absorption profile. The
life-time of the excited optical state $T_1=164\;\mu\text{s}$ and
the homogeneous line-width is 2.4~kHz. These values correspond to
crystals at liquid helium temperature. The total inhomogeneous
broadening of the optical transition is of the order of 4~GHz, so
that the Pr ion density within a 100~kHz frequency interval is
roughly equal to $2\times 10^{20}\;\text{m}^{-3}$ for crystals
with Pr doping concentration of 0.05\% ($9\times
10^{24}\;\text{m}^{-3}$). The principal point here is the
possibility of changing this density from that maximum value down
to zero when preparing the narrow absorption peaks. This can be a
convenient way to investigate a transition from incoherent
spontaneous emission to superradiance in a given crystal by
adjusting the superradiant life-time $\tau_R$ to the desirable
value. Taking the wavelength of the optical transition
$\lambda=606$~nm, the length of a sample $L_z=5$~mm, the diameter
of a focal spot $100\;\mu\text{m}$, we obtain $\tau_E=17$~ps,
$\mu=0.5\times 10^{-5}$ and $\tau_R=3.7$~ns. The last value is too
short because the pit width is equal to 10~MHz so that the
duration of superradiant decay should be longer than 20~ns. If we
decrease the ion density down to $2\times 10^{19}\;\text{m}^{-3}$,
we obtain obviously $\tau_R=37$~ns, which may be considered as the
shortest possible value. On the other hand, the upper limit for
the superradiant decay time is given by the inhomogeneous
life-time $T_2^\ast=(\pi\Gamma_{\text{inh}})^{-1}$, which in this
case is equal $3\;\mu\text{s}$. With the parameters mentioned
above we obtain $\tau_R=1.9\;\mu\text{s}$ for an ion density of
$4\times 10^{17}\;\text{m}^{-3}$, which may be considered as a
lower bound. Thus, we can conclude that the observation of optical
superradiance in the time scale $10^2-10^3$~ns is possible. It
should be noted that by decreasing the inhomogeneous broadening of
the absorption peak down to 10~kHz we obtain an inhomogeneous
life-time of $T_2^\ast=30\;\mu\text{s}$. Therefore, in principle,
it is possible to make $\tau_R$ of the order of several
microseconds, which may be very useful for initial investigations
of subradiant states in such a system. Since dipole moments of
optical transitions are rather small (for example, 0.0078~D for
the transition ${}^3H_4(1)-{}^1D_2(1)$ \cite{NRKKS_2004}), such
values of $\tau_R$ allow the use of $2\pi$-pulses with a duration
up to $100-200$~ns, thereby reducing the intensity of excitation
pulses.

\subsection{Active preparation of subradiant states}

\begin{figure}
\includegraphics[width=7cm]{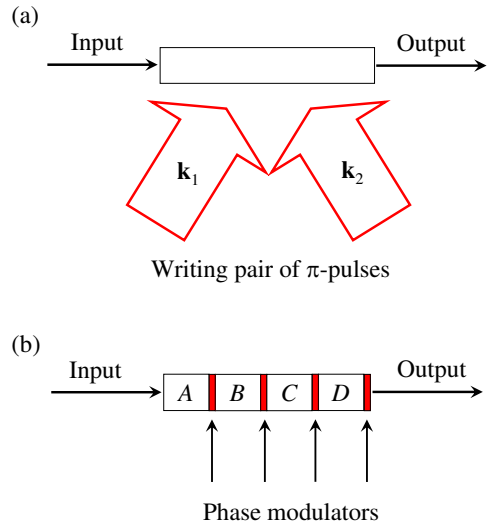}
\caption{\label{fig:scheme2} (Color online) Scheme for the
preparation of subradiant states in an extended atomic ensemble
using (a) pairs of non-collinear and non-simultaneous coherent
$\pi$-pulses and (b) controllable bi-phase modulators in the case of
four spatial parts.}
\end{figure}

The active preparation implies that the phases of atomic states are
modulated by excitation pulses, such as the $2\pi$-pulses considered
above. A scheme for preparing orthogonal subradiant states that may
be simpler from an experimental point of view is shown in
Fig.~\ref{fig:scheme2}(a). In this case two non-collinear
$\pi$-pulses are applied in sequence instead of a single
$2\pi$-pulse. Suppose that $\vec{k}_1$ and $\vec{k}_2$ are wave
vectors of the first and second pulses, respectively, in a pair and
the vector $\vec{k}_1-\vec{k}_2$ is directed along the sample $z$
axis.  After the excitation the collective atomic state
$\ket{1_{\vec{q}}}$ becomes
$-\ket{1_{\vec{q}+\vec{k}_1-\vec{k}_2}}$. It follows that such an
excitation leads to an added phase shift while propagating through
the medium. If this phase shift is a multiple of $2\pi$, i.e.
$|\vec{k}_1-\vec{k}_2|=2\pi m L_z^{-1}$, where $m$ is an integer, a
subradiant state is obtained. The rate of collective spontaneous
emission into the longitudinal mode proves to be zero due to
destructive interference between various spatial parts. (We propose
here that $m$ is small, i.e. $|\vec{k}_1-\vec{k}_2|\ll \omega_0/c$.)
Different values of $m$ correspond to different orthogonal
subradiant states. In order to convert the system back from the
subradiant state into the superradiant state the same pulses need to
be applied in reverse order, i.e. the wave vectors of the first and
second read-out pulses should be $\vec{k}_2$ and $\vec{k}_1$
respectively.

Compared to the excitation by $2\pi$-pulses the advantage of such an
excitation is that the second $\pi$-pulse in a pair can transfer an
atomic state from the excited level to another long-lived ground
level. In this case the storage time may be much longer than the
life-time of the excited states.

\subsection{Passive preparation of subradiant states}

On the other hand, it is not necessary to create a phase shift using
coherent excitation pulses. We can create it by controllable phase
modulators inserted into the atomic ensemble as shown in
Fig.~\ref{fig:scheme2}(b). Suppose that each of the modulators may
be in one of two possible states: an off-state, when it does not
create a phase shift, and an on-state, when a $\pi$ phase shift is
created. Now, we turn on and turn off these bi-phase modulators in a
definite order instead of application of $2\pi$-pulses. A subradiant
state is obtained if destructive interference of radiation emitted
from various spatial parts occurs at the exit of the sample and, of
course, if the emission process remains collective for the whole
system. In the case of four spatial parts, if we want to obtain the
sequence of subradiant states which corresponds to the Hadamard
matrix (\ref{Had1}), we should switch the phase modulators in the
following order: $ABCD\to \bar{A}B\bar{C}D\to A\bar{B}C\bar{D}$,
where $A$ ($\bar{A}$) means that the modulator after the part $A$ is
turned on (turned off) and so on. The read out without time reversal
is described by the following scheme: $\bar{A}\bar{B}\bar{C}{D}\to
ABC\bar{D}\to \bar{A}{B}\bar{C}\bar{D}$.

Compared to the active preparation scheme such a passive preparation
scheme may have several advantages. Single-photon wave packets
propagating in optical fibers or wave guides can be stored and
reconstructed without using transverse coherent excitation. The only
condition is that photon propagation time through the whole system
should be shorter than $\tau_R$. Further, an atomic ensemble (or
each optical center) may be placed in a confined space or artificial
structure like a photonic crystal such that spontaneous emission
into transverse modes is inhibited. In this case controlling the
emission and creating subradiant states for the longitudinal modes
may enable storage of qubit states in decoherence free substates in
the ensemble during times which may be longer than the excited state
life-time of a free atom.

\section{Conclusion}
It is shown that single-photon wave packets with arbitrary pulse
shape resonant with a transition in an ensemble of atoms may be
recorded and reconstructed in a regime of optical subradiance also
in the absence of inhomogeneous broadening of the resonant
transition. Homogeneously broadened lines are more prone to
superradiance than inhomogeneously broadened lines but the
technique described here allows the writing and reconstruction of
time-bin qubits in an optically dense medium also when the
time-bin qubit duration exceeds the superradiant life-time for the
ensemble. This approach could therefore open a new window in the
search of materials for quantum memories.

\begin{acknowledgements}
A.K. thanks the Division of Atomic Physics Lund Institute of
Technology for hospitality. This work was supported by the Russian
Foundation for Basic Research (Grant No. 04-02-17082à), the
Program of the Presidium of RAS 'Quantum macrophysics' and the
European Commission within the integrated project QAP under the
IST directorate.
\end{acknowledgements}

\appendix
\section{}\label{app:A}
To solve the equations (\ref{Eq1}) and (\ref{Eq2}) we use the
Laplace transformation
\begin{align}
\bar{y}(z)&=\int_0^\infty e^{-zt}y(t)\,dt,\\
z\bar{y}(z)-y(0)&=\int_0^\infty e^{-zt}\dot{y}(t)\,dt
\end{align}
and obtain
\begin{eqnarray}
z\bar{f}_{\vec{k},s}(z)-f_{\vec{k},s}(0)&=
&-ig_{\vec{k},s}\sqrt{N}\phi(\vec{q}-\vec{k})\nonumber\\&
&\times\bar{c}(z+i(\omega_0-\omega)),\\
z\bar{c}(z)-c(0)&=
&-i\sum_{\vec{k},s}g_{\vec{k},s}^\ast\sqrt{N}
\phi^\ast(\vec{q}-\vec{k})\nonumber\\&
&\times\bar{f}_{\vec{k},s}(z-i(\omega_0-\omega)).
\end{eqnarray}
For the atomic excitation amplitude we find
\begin{equation}
\bar{c}(z)=\frac{c(0)-i\sum_{\vec{k},s}\dfrac{g_{\vec{k},s}^\ast\sqrt{N}
\phi^\ast(\vec{q}-\vec{k})f_{\vec{k},s}(0)}{z-i(\omega_0-\omega)}}
{z+\sum_{\vec{k},s}
\dfrac{|g_{\vec{k},s}|^2N|\phi(\vec{q}-\vec{k})|^2}{z-i(\omega_0-\omega)}}.
\end{equation}
In the Born-Markov approximation we can evaluate the sum in the
denominator by first converting it to an integral
\begin{equation}
\sum_{\vec{k}}\to\frac{L^3}{(2\pi)^3c^3}\sum_0^\infty\omega^2\,d\omega\int
d\Omega_{\vec{k}}
\end{equation}
and then employing the identity \[\lim_{z\to
0}\frac{1}{z-i(\omega_0-\omega)}=\pi\,\delta(\omega_0-\omega)
+i\frac{P}{\omega_0-\omega}\] with $P$ denoting the principal value.
As a result we obtain
\begin{equation}\label{G}
\sum_{\vec{k},s}\frac{|g_{\vec{k},s}|^2N|\phi(\vec{q}-\vec{k})|^2}
{z-i(\omega_0-\omega)}=\Gamma +i\Delta,
\end{equation}
where
\[\Gamma=\frac{d^2N\mu\omega_0^3}{6\pi\hbar\varepsilon_0c^3}\]
is the collective atomic linewidth and $\Delta$ is the collective
frequency shift, which may be neglected. Here $\mu=3\lambda^2(8\pi
S)^{-1}$ is a geometrical factor \cite{RE_1971}, which describes the
result of the integration
\begin{equation}
\int
d\Omega_{\vec{k}}\sum_s(\vec{d}\cdot\vec{\varepsilon}_{\vec{k},s})^2
\phi^2(\vec{q}-\vec{k})=\frac{8\pi}{3}\mu d^2
\end{equation}
for a pencil-shaped cylindrical volume with the cross-section $S$, a
vector $\vec{q}$ lying along the axis of the cylinder and dipole
moments oriented perpendicular to the axis. With the help of
(\ref{G}) the inverse Laplace transformation yields
\begin{eqnarray}\label{App_c}
c(t)=c(0){\,e}^{-\Gamma
t}&-&i\sum_{\vec{k},s}\frac{g_{\vec{k},s}^\ast\sqrt{N}
\phi^\ast(\vec{q}-\vec{k})f_{\vec{k},s}(0)}
{\Gamma+i(\omega_0-\omega)}\nonumber\\&
&\times\left[e^{i(\omega_0-\omega)t}-{e}^{-\Gamma t}\right].
\end{eqnarray}
It is convenient to define the incoming single-photon wave packet
shape by the dimensionless photon density in the origin of the
reference frame
\begin{equation}
F_\text{in}(t)=\left(\frac{L_xL_yL_z}{L^3}\right)^{1/2}\sum_{\vec{k},s}
f_{\vec{k},s}(0){\,e}^{i(\omega_0-\omega)t}.
\end{equation}
Then with the help of approximations (iv-vi) the equation
(\ref{App_c}) may be rewritten as
\begin{eqnarray}
c(t)&=&c(0){\,e}^{- t/2\tau_R}\nonumber\\&
&-\frac{1}{\sqrt{\tau_R\tau_E}}\Bigg[\int_0^\infty d\tau
F_\text{in}(t-\tau){\,e}^{-\tau/2\tau_R}\nonumber\\&
&-{e}^{-t/\tau_R}\int_0^\infty d\tau
F_\text{in}(-\tau){\,e}^{-\tau/2\tau_R}\Bigg],
\end{eqnarray}
where $\tau_R=(2\Gamma)^{-1}$ is the superradiant life-time. The
second term in the brackets obviously equal to zero since the
interaction between the atomic system and the radiation field is
turned on at the moment $t = 0$. Thus we obtain the solution
Eq.~(\ref{Solution_0a}).

In order to obtain the solution for $F(t)$ we integrate equation
(\ref{Eq1}), which gives
\begin{equation}
f_{\vec{k},s}(t)=f_{\vec{k},s}(0)-ig_{\vec{k},s}\sqrt{N}
\phi(\vec{q}-\vec{k})\int_0^t c(t'){\,e}^{-i(\omega_0-\omega)t'}dt'.
\end{equation}
Then we substitute this result into the general formula
\begin{equation}
F(t)=\left(\frac{L_xL_yL_z}{L^3}\right)^{1/2}
\sum_{\vec{k},s}f_{\vec{k},s}(t){\,e}^{i(\omega_0-\omega)t},
\end{equation}
and take into account that
\begin{equation}
\int_0^t
dt'{\,e}^{i(\omega_0-\omega)t'}=\pi\,\delta(\omega_0-\omega)
+i\frac{P}{\omega_0-\omega},
\end{equation}
where the imaginary part may be neglected, and
\begin{equation}
\int
d\Omega_{\vec{k}}\sum_s(\vec{d}\cdot\vec{\varepsilon}_{\vec{k},s})
\phi(\vec{q}-\vec{k})=\frac{16\pi}{3}\mu
d.
\end{equation}
As a result after some calculations we obtain
Eq.~(\ref{Solution_0b}).

\section{}\label{app:B}
Let us consider a three-level atom interacting with a two-mode
electromagnetic field. The Hamiltonian of the system in the
interaction picture is
\begin{equation}
H=\hbar(g_aR_{10}a+g_bR_{21}b)+\text{H.c.},
\end{equation}
where $R_{ij}=\ket{i}\bra{j}$ is the atomic transition operator
($i$, $j=0,1,2$), $a$, $b$ are the photon annihilation operators for
modes with frequency $\omega_{01}$ and $\omega_{12}$, respectively,
$g_a$, $g_b$ are atomic-field coupling constants. In the photon
number representation an arbitrary pure state of the system may be
written as
\begin{equation}
\ket{\psi(t)}=\sum_{i=0}^2\sum_{n_a,n_b=0}^\infty
c_{i,n_a,n_b}\ket{i}\ket{n_a,n_b},
\end{equation}
and its evolution in time is given by
\begin{equation}
\ket{\psi(t)}=\exp(-i\hbar^{-1}Ht)\ket{\psi(0)}=U(t)\ket{\psi(0)}.
\end{equation}
The evolution operator $U(t)$ may be written as \cite{P_2001}
\begin{equation}\label{U}
U(t)=\sum_{i,j=0}^2\phi_{ij}(t)R_{ij},
\end{equation}
where
\begin{eqnarray*}
\phi_{00}&=&1+g_a^2a^\dag\Gamma^{-2}[\cos(\Gamma t)-1]a,\\
\phi_{01}&=&-ig_aa^\dag\Gamma^{-1}\sin(\Gamma t),\\
\phi_{02}&=&g_a g_b a^\dag\Gamma^{-2}[\cos(\Gamma t)-1]b^\dag,\\
\phi_{11}&=&\cos(\Gamma t),\\
\phi_{12}&=&-i\Gamma^{-1}\sin(\Gamma t)g_b b^\dag,\\
\phi_{22}&=&1+g_b^2b\Gamma^{-2}[\cos(\Gamma t)-1]b^\dag,
\end{eqnarray*}
$\phi_{ij}=-\phi_{ji}^\dag$ and $\Gamma=(g_a^2 aa^\dag+ g_b^2 b^\dag
b)^{1/2}$.

Let the initial state of the system be
\begin{equation}
\ket{\psi(0)}=c_0\ket{0}\ket{1,\alpha}+c_1\ket{1}\ket{0,\alpha},\quad
|c_0|^2+|c_1|^2=1,
\end{equation}
where
$\ket{\alpha}=\exp(-\frac{1}{2}|\alpha|^2)
\sum_{n_b=0}^\infty\alpha^{n_b}(n_b!)^{-1/2}\ket{n_b}$
is a coherent state with an amplitude $\alpha$
($|\alpha|^2=\avr{n_b}$). This means that at the moment $t=0$ the
atom being in an entangled state
$c_0\ket{0}\ket{1,0}+c_1\ket{1}\ket{0,0}$ with the mode
$\omega_{01}$ starts to interact with a coherent state of the mode
$\omega_{12}$. With the help of equation (\ref{U}) we find
\begin{eqnarray*}
\ket{\psi(t)}&=&\Big[c_0\{1+g_a^2\Omega^{-2}[\cos(\Omega t)-1]\}\\&
&-ic_1g_a\Omega^{-1}\sin(\Omega t)\Big]\ket{0}\ket{1,\alpha}\\& &+
\Big[-ic_0g_a\Omega^{-1}\sin(\Omega t)+c_1\cos(\Omega
t)\Big]\ket{1}\ket{0,\alpha}\\& &+\Big[c_0\alpha g_a g_b
\Omega^{-2}[\cos(\Omega t)-1]\\& &-ic_1\alpha
g_b\Omega^{-1}\sin(\Omega t)\Big]\ket{2}\ket{0\alpha},
\end{eqnarray*}
where $\Omega=[g_a^2+(\Omega_R/2)^2]^{1/2}$ and
$\Omega_R=2g_b|\alpha|$ is the Rabi frequency of the coherent
field. Consequently at time $t_p=\pi\Omega^{-1}$ we have
\begin{eqnarray}
\ket{\psi(t_p)}&=&c_0(1-2g_a^2\Omega^{-2})\ket{0}\ket{1,\alpha}\nonumber\\
& &-c_1\ket{1}\ket{0,\alpha}\nonumber\\& &-2c_0\alpha g_a
g_b\Omega^{-2}\ket{2}\ket{0\alpha}.
\end{eqnarray}
Thus, the atom proves to be in the upper state $\ket{2}$ with the
probability
\begin{equation}
p=|c_0|^2\left(\frac{g_a\Omega_R}{g_a^2+(\Omega_R/2)^2}\right)^2
\end{equation}
and returns to the initial state $\ket{\psi(0)}$  (having a $\pi$
phase shift for the state $\ket{1}$ amplitude) with the
probability $1-p$. For high Rabi frequencies $\Omega_R\gg g_a$ the
probability for a successful phase shift operation, $1-p$, becomes
close to 1.

\end{document}